\documentclass[12pt]{article}
\usepackage{amsmath,amsthm,amssymb,euscript,epsf,epsfig,cite,bbm}
\usepackage{array}
\usepackage{fancybox}
\usepackage[usenames]{color}
\usepackage{amsfonts,bm}
\usepackage[stable]{footmisc}
\usepackage{graphicx}
\usepackage{enumerate}

\usepackage[
      colorlinks=true,
      linkcolor=blue,
      urlcolor=blue,
      linktocpage=true,
      filecolor=blue,
      citecolor=blue,
      pdfstartview=FitH,
      pdftitle={Beyond Logarithmic Corrections to the Cardy Formula},
      pdfauthor={Farhang Loran, Mohammad M. Sheikh-Jabbari and Massimiliano Vincon},
      pdfsubject={2D Conformal Field Theory},
      pdfkeywords={Black Holes in String Theory, Supergravity Models, CFT},
      pdfpagemode=UseNone,
      bookmarksopen=true
      ]{hyperref}
\usepackage[all]{hypcap}
\makeatletter \@addtoreset{equation}{section}

\makeatletter\renewcommand\section{\@startsection {section}{1}{\z@}%
                                   {-3.5ex \@plus -1ex \@minus -.2ex}
                                   {2.3ex \@plus.2ex}%
                                   {\normalfont\large\bfseries}}
\renewcommand\subsection{\@startsection{subsection}{2}{\z@}%
                                     {-3.25ex\@plus -1ex \@minus -.2ex}%
                                     {1.5ex \@plus .2ex}%
                                     {\normalfont\bfseries}}

\makeatletter \@addtoreset{equation}{section}

\makeatletter\renewcommand\section{\@startsection {section}{1}{\z@}%
                                   {-3.5ex \@plus -1ex \@minus -.2ex}
                                   {2.3ex \@plus.2ex}%
                                   {\normalfont\large\bfseries}}
\renewcommand\subsection{\@startsection{subsection}{2}{\z@}%
                                     {-3.25ex\@plus -1ex \@minus -.2ex}%
                                     {1.5ex \@plus .2ex}%
                                     {\normalfont\bfseries}}

 \newcommand{\be}{\begin{equation}}
 \newcommand{\ee}{\end{equation}}
 \newcommand{\bea}{\begin{eqnarray}}
 \newcommand{\eea}{\end{eqnarray}}
 \newcommand{\nn}{\nonumber}

\def\cft2{CFT$_2$ }

\def\Tr{\mathrm{Tr}}

\def\tc{{\tilde c}}

\def\z2{$\mathbb{Z}_2$ }

\newcommand{\bse}{\begin{subequations}}
\newcommand{\ese}{\end{subequations}}

\begin{document}

\begin{titlepage}
\thispagestyle{empty}
\begin{flushright}
{
{\tt arXiv:1010.3561} \\
IPM/P-2010/041  \\
}\end{flushright} \vspace{10 mm}
\begin{center}
    \font\titlerm=cmr10 scaled\magstep4
    \font\titlei=cmmi10 scaled\magstep4
    \font\titleis=cmmi7 scaled\magstep4
     {\Large \bf Beyond Logarithmic Corrections to Cardy Formula}

 \vspace{1.0cm}
    \noindent{  \textbf{{Farhang Loran$^{a,}$\footnote{\tt loran@cc.iut.ac.ir}}}},
    { \bf  M. M. Sheikh-Jabbari$^{b,}$\footnote{\tt jabbari@theory.ipm.ac.ir} and
    Massimiliano Vincon$^{b,}$\footnote{\tt m.vincon@ipm.ir}}
    \vspace{0.8cm}

   $^a${\it Department of  Physics, Isfahan University of Technology,
  Isfahan 84156-83111,  Iran}\\
\vskip 2mm
  $^b${\it  School of Physics,
 Institute for Research in Fundamental Sciences (IPM),\\ P.O.Box
 19395-5531, Tehran, Iran}\\

  \end{center}


\begin{abstract}
{\noindent{As shown by Cardy \cite{Cardy}, modular invariance of the
partition function of a given unitary non-singular $2d$ CFT  with
left and right central charges $c_L$ and $c_R$, implies that the
density of states in a microcanonical ensemble, at excitations
$\Delta$ and $\bar\Delta$ and in the {saddle point approximation},
is $\rho_0(\Delta,\bar\Delta; c_L, c_R)=c_L
\exp(2\pi\sqrt{{c_L\Delta}/{6}}\ ) \cdot c_R
\exp(2\pi\sqrt{{c_R\bar\Delta}/{6}}\ )$.  In this paper, we extend
Cardy's analysis and show that in the saddle point approximation and
up to contributions which are {\em exponentially suppressed}
compared to the leading Cardy's result, the density of states takes
the form $\rho(\Delta,\bar\Delta; c_L, c_R)= f(c_L\Delta)
f(c_R\bar\Delta)\rho_0(\Delta,\bar\Delta; c_L, c_R)$, for a function
$f(x)$ which we specify. In particular, we show that  \textrm{(i)}
$\rho (\Delta,\bar\Delta; c_L, c_R)$ is the product of contributions
of left and right movers and hence, to this approximation, the
partition function of any modular invariant, non-singular unitary
$2d$ CFT is holomorphically factorizable and \textrm{(ii)} $\rho
(\Delta,\bar\Delta; c_L, c_R)/(c_Lc_R)$ is only a function of
$c_L\Delta$ and $c_R\bar\Delta$. In addition, treating
$\rho(\Delta,\bar\Delta; c_L, c_R)$ as the density of states of a
microcanonical ensemble, we compute the entropy of the system in the
canonical counterpart and show that the function $f(x)$ is such that
the canonical entropy, up to exponentially suppressed contributions,
is simply given by the Cardy's result $\ln \rho_0(\Delta,\bar\Delta;
c_L, c_R)$.}}

  \end{abstract}

\end{titlepage}

\tableofcontents

\section{Introduction }\label{intro}

\setcounter{footnote}{0}

Conformal symmetry as a natural extension of the Poincar\'e symmetry
has long been noted and extensively studied in the formulation of
quantum field theories. In two dimensions, the conformal algebra becomes infinite
dimensional and admits a central extension $c$, the Virasoro algebra
\be\label{Vir-algebra}%
[L_m, L_n]=(m-n) L_{m+n}+\frac{c}{12} n(n^2-1) \delta_{m+n}\, .%
\ee%
As put forward and stressed in the classic work of
Belavin-Polyakov-Zamolodchikov \cite{BPZ}, conformal invariance in two dimensions can be powerful enough to render the conformal field theory
(CFT) solvable. Being infinite dimensional, unitary irreducible
representations (irreps) of two-dimensional conformal algebra are also infinite
dimensional. Each irrep is composed of a highest weight
state (corresponding to a primary operator) and its descendants,
obtained by the action of $L_n$, with $n<0$, on the highest weight state. A
CFT is then specified by determining the set of $L_n$'s and its
primary fields together with their operator product expansions (OPEs). A detailed analysis of $2d$ CFTs and references to the original papers may be found in \cite{CFT-books-1,CFT-books-2}.

In a unitary  CFT all the highest weight states and hence all the
states of the theory, have a non-negative norm. This implies that
both the conformal weights $\Delta$ (eigenvalues of primary states
under $L_0$) of all primary operators and the central charge $c$ are
non-negative. In this work we will only be concerned with unitary
$2d$ CFTs.

A classification of $2d$ CFTs may be arranged based on the value of
the central charge $c$ and the spectrum of the conformal weights of
primary fields. Except for some specific cases, such as CFTs with a
finite number of primary operators (the minimal models) and WZNW
models, $2d$ CFTs are not proven to be solvable \cite{CFT-books-1}.
In such cases one would like to know if  conformal invariance
teaches us anything about the spectrum and degeneracy of  states of
the CFT.

A crude though useful notion is whether the spectrum of the primary
operators of the CFT is continuous (for singular CFTs) or discrete
(for non-singular CFTs) as in the case of rational conformal field theories or their subclass of minimal models. String
worldsheet theory on a $D$-dimensional flat target space,
\emph{i.e.} a system of $D$ free noncompact bosons, is a trivial
example of singular CFT. This is due to the fact that as a result of
translation symmetry of the target space, the states and their
conformal weights are also labeled by the center of mass momentum
which  is a continuous parameter. Nonetheless, one may compactify
the target space on $T^D$ and make the spectrum discrete, thus
effectively turning it into a ``non-singular CFT''  with a discrete spectrum for primary operators
\cite{Polchinski-book}. In the pursuit of our investigation, we
shall only consider non-singular unitary CFTs or CFTs which can be
made non-singular by a ``regularization'' like the example of
compactified string theory mentioned above. In section \ref{Carlip-Extension-section} we will give a more precise definition of what we mean by ``non-singular CFT''. 

$2d$ CFTs may be defined on two-dimensional surfaces of various topology, in
particular a torus $T^{2}$. Although not necessary, in many physically
relevant cases one needs to make sure that the theory is modular
invariant.\footnote{Modular invariance becomes a necessity if the $2d$ CFT in question is going to be viewed as worldsheet description of a string theory \cite{Polchinski-book}.}
 For the case at hand, where the torus $T^2$ is specified by the
complex structure (or modular parameter) $\tau$,  modular
transformations are elements of PSL$(2, \mathbbm {Z)}=$SL$(2,\mathbbm {Z})/\mathbbm {Z}_2$  which act
on $\tau$ (and simultaneously on $\bar\tau$) as
\be%
\tau\to \ \tau'=\frac{a\tau+b}{c\tau+d}\ ,\qquad \left(\begin{array}{ll} a & b \\
c&
d\end{array}\right)\in \mathrm{PSL}(2,\mathbbm {Z})\, . %
\ee%
Modular invariance then demands  that ${\cal Z}(\tau,\bar\tau)={\cal
Z}(\tau',\bar\tau')$, where ${\cal Z}$ is the partition function of
the CFT on the torus. In a seminal work \cite{Cardy}, Cardy realized
and emphasized on the role of modular transformations and the
restrictions modular invariance imposes on the content of primary
operators of the CFT.\footnote{PSL$(2,\mathbbm {Z})$ can be
generated by the action of two independent elements, usually known
as $S$- and $T$-transformations,   corresponding to
$(a,b,c,d)=(0,-1,1,0)$ and $(1,1,0,1)$, respectively. Our analysis
here, following that of Cardy's, is based on exploiting
$S$-transformation. In an interesting paper,  implications of
$T$-transformation for the spectrum of modular invariant, unitary
CFTs have also been discussed \cite{Hellerman}. } In the case of
minimal models, modular invariance restricts the central charge to
be a rational number less than one, explicitly
$c=1-\frac{6(p-q)^2}{pq}$, where $p$ and $q$ are co-prime integers
\cite{CFT-books-1}. Moreover, if we also demand unitarity, the
central charge is further restricted to $c=1-\frac{6}{(q+2)(q+3)}$,
where $q$ is a positive integer. The theory of free fermions on a
$T^2$ is  another example of modular invariant theory, if both
periodic and anti-periodic boundary conditions are allowed.

Invariance under modular transformations  has been employed in unitary non-singular CFTs to specify the density
of states at a given conformal weight $\Delta$. It was shown by
Cardy  \cite{Cardy} that in the {\em saddle point} approximation and for $\Delta\gg
c$, the density of states $\rho(\Delta)$ grows exponentially as $\sqrt{\Delta}$
\be \rho(\Delta)_{\textrm{Cardy}}\sim \exp{\left(2\pi
\sqrt{\frac{c\Delta}{6}}\right)}\, . \ee Cardy's analysis may be
repeated for a general $2d$ CFT with left and right central charges
$c_L$ and $c_R$, at left and
right excitations $\Delta$ and $\bar \Delta$ respectively, yielding%
\be\label{Cardy-density}%
\rho(\Delta,\bar\Delta)_{\textrm{Cardy}}\sim
e^{2\pi\sqrt{\frac{c_L\Delta}{6}}}\cdot
e^{2\pi\sqrt{\frac{c_R\bar\Delta}{6}}}\, . %
\ee%

One can make two observations from the above Cardy formula: 1) The
density of states in the saddle point approximation is the product
of those in the left and right sectors or, equivalently, the
partition function is the product of a holomorphic and an
anti-holomorphic function which may be thought of as partition
functions of the left and right sectors, respectively. In other
words, in the Cardy's saddle point approximation, modular invariance
implies that \emph{any unitary non-singular $2d$ CFT is
holomorphically factorizable}.\footnote{As we discuss in section
\ref{Discussion-Section}, holomorphic factorizability of a $2d$ CFT
may also  be relevant in the context of AdS/CFT and $3d$ quantum
gravity theories \cite{Witten:2007kt}.} 2) The density of states is
{\em only} a function of $c\cdot\Delta$ and not of their arbitrary
combination.

In what follows we  expand upon Cardy's considerations and by
exploiting modular invariance further we  explore the validity of
the above two observations beyond Cardy's first-order saddle point
approximation. In particular, we  show that for any modular
invariant, unitary and non-singular $2d$ CFT the above two
observations remain valid up to {\em exponentially suppressed
contributions}, within {\em saddle point approximation}. We will
discuss the precise meaning of these two expressions in sections
\ref{Carlip-Extension-section} and \ref{Simeon-method-section}. We
should also mention that the expectation of validity of the previous
observations is compatible with the results of Carlip \cite{Carlip},
where first order in perturbation theory beyond the Cardy formula
was studied. Our analysis here is then an all-orders extension of
Carlip's results.

The rest of this paper is organized as follows. In section
\ref{Carlip-Extension-section}, we  revisit the analysis of
\cite{Carlip} more closely and extend it to all orders in
perturbation theory. In this section, we have used the more formal
language of path integral. Although a clean method in the sense of
producing an all-orders result in the saddle point approximation,
the path integral approach may obscure the physical intuition and
some of the assumptions we have made in the course of the
computation. Therefore, in section \ref{Simeon-method-section} we
present another method  based on perturbative, order by order
expansion in the saddle point approximation, reproducing the results
of section \ref{Carlip-Extension-section} in a Taylor series
expansion. In section \ref{micro-vs-canonic-section},  using the
expression for the microcanonical density of states  obtained in
sections \ref{Carlip-Extension-section} and
\ref{Simeon-method-section}, we compute the canonical partition
function of the theory as well as its canonical entropy and compare
the latter to the microcanonical results. In section
\ref{Discussion-Section}, besides reviewing our results, we draw our
conclusions and discuss the implications of our findings if the $2d$
CFT in consideration is regarded as the CFT dual to gravity on an
AdS$_3$ background. In two appendices  we have collected the details
of some computations, the results of which have been used in sections
\ref{Carlip-Extension-section} and \ref{Simeon-method-section}.

\section{Extension of Carlip's Formulation}\label{Carlip-Extension-section}

For a generic (Euclidean) unitary $2d$ CFT with left and right
central charges $c_L$ and $c_R$ on a torus with complex structure $\tau=\tau_1+i\tau_2$, the partition function is defined as%
\be
\label{partition-function-tau}
Z(\tau,\bar\tau)=\Tr \left(e^{2\pi i\tau L_0}\ e^{-2\pi i \bar\tau \bar L_0}\right)=\sum_{\Delta,\bar \Delta=0}
\rho(\Delta,\bar\Delta) e^{2\pi i\tau \Delta}\ e^{-2\pi i \bar\tau \bar \Delta} \, ,
\ee%
where $\bar\tau=\tau_1-i\tau_2$, $\Delta$ and $\bar\Delta$, which we assume to take non-negative values, are  respectively the spectrum of $L_0$ and $\bar L_0$, and $\rho(\Delta,\bar\Delta)$ is the density of states with left and right energies $\Delta$ and $\bar\Delta$. Unitarity of the CFT implies that $\rho$ is positive definite (negative norm states would have contributed to the partition function with negative $\rho$). In general, $Z(\tau,\bar\tau)$ is neither a holomorphic function of $\tau$ nor holomorphically factorizable, where $Z(\tau,\bar\tau)$ factorizes to holomorphic and
antiholomorphic parts, and the modular invariant partition function ${\cal Z}$ is related to $Z$ as
\be
{\cal Z}(\tau,\bar\tau)=e^{-\frac{2\pi i}{24}(\tau c_L-\bar\tau c_R)} Z(\tau,\bar\tau)\, .
\ee

If one dealt with a chiral CFT, or  the trace in the partition
function were only over the chiral sector of the Hilbert space of
the theory ({\it i.e}. over the $\bar L_0=0$ sector), one could make
use of powerful analytic (more precisely meromorphic) functional properties in combination with
modular invariance and express the partition function only in terms
of its {\em polar part} and hence arrive at the Rademacher expansion
which completely specifies the partition function
\cite{Farey-tail-1,Farey-tail-2}. We  note that the Rademacher
expansion can also be applied to holomorphically factorizable CFTs
\cite{Maloney:2007ud}. However, these techniques are not available
in general and one may wonder how far one can go relying only on
modular invariance, the question we explore below.

In his influential  paper \cite{Cardy}, Cardy took the first steps
into this direction, showing that for any  matching
``high-and-low-temperature'' expansion of the partition function one
may compute the density of states for any given $2d$ CFT. Cardy
formula, among other things, implies that around the saddle point
the partition function is the product of its holomorphic and
anti-holomorphic parts. This last statement is equivalent to the
fact that the  total entropy of the $2d$ CFT is the sum of the
entropy of left and right sectors. In \cite{Carlip}, it was shown
how one may extend Cardy's saddle point analysis and compute the
leading logarithmic correction to the Cardy formula. Remarkably, it
was observed how the addition of logarithmic corrections to the
partition function would not spoil its holomorphic factorizability.
In this section, we wish to broaden Carlip's method and show that
indeed this {\em approximate holomorphic factorizability} holds true
to any perturbative order around the saddle point in the $1/\tau$
expansion, up to exponentially suppressed corrections. We should
stress that in order to argue for our case we only resort to
\textrm{(i)} modular invariance \textrm{(ii)} unitarity of the
theory and \textrm{(iii)} that the CFT be non-singular, a term which will be defined momentarily. These are
indeed very mild assumptions in the sense that they are satisfied by
many CFTs.

We begin our analysis by recalling that the density of
states at energies $\Delta$ and $\bar\Delta$ can be computed using contour
 integrals in the two corresponding complex planes of $q=e^{2\pi i\tau}$ and $\bar
 q=e^{-2\pi i\bar \tau}$
 \be \label{rho-def}
 \rho(\Delta,\bar\Delta)=\frac{1}{(2\pi i)^2 }\int dq \, d\bar q\
 \frac{1}{q^{\Delta+1}}
 \frac{1}{\bar q^{\bar\Delta+1}} Z(q, \bar q) \, .
 \ee%
Note that the variables $q$ and $\bar q$ are to be treated as
independent variables and not as complex conjugate of each other.
Equation \eqref{rho-def} is the usual Laplace transform to move from
canonical to microcanonical ensemble, and thus we regard
$\rho(\Delta,\bar\Delta)$ as the \emph{microcanonical} density of
states. Let us also remark  that for this Laplace transform to be
well-defined, or stated otherwise,  for  \eqref{rho-def} to follow
from \eqref{partition-function-tau}, one should assume that the
spectrum of the CFT is labeled by a non-negative integer $n$ such
that $\Delta_n-\Delta_0$ is a non-negative integer, where $\Delta_0$
is the ground state energy, and similarly for the anti-holomorphic
sector. This requirement, explicitly having discrete and integer-valued $\Delta_n-\Delta_0$, is what defines a non-singular CFT.
\footnote{ Note that  with this definition  any rational CFT \cite{CFT-books-1} is non-singular while the converse is not necessarily true. See footnote 6 for further comments on this point. }

 Next, we use modular invariance to relate $Z(\tau,\bar\tau)$ to
 $Z(-1/\tau,-1/\bar\tau)$. Doing so yields
 \be \label{maro22}
 Z(\tau,\bar\tau)=e^{\frac{2\pi i\tau
c_L}{24}} e^{\frac{-2\pi i\bar\tau c_R}{24}}\ \left(\sum_{n,\bar
n=0}\ \rho(\Delta_n,\bar\Delta_{\bar n})\ e^{\frac{-2\pi
i(\Delta_n-\frac{c_L}{24})}{\tau}} e^{\frac{2\pi
i(\bar\Delta_n-\frac{c_R}{24})}{\bar\tau}}\right)\, .
\ee%
 Inserting \eqref{maro22} into \eqref{rho-def} we obtain
\be\label{rho-integral}%
\rho(\Delta,\bar\Delta)=\sum_{n,\bar n=0}\rho(\Delta_n,\bar\Delta_{\bar n})\ I(a, b_n)  I(\bar a,\bar b_n)\, ,
\ee
where%
\be%
 \label{a-sign}
 I(a,b_n) =- \int_{0}^{i\infty(+)} d\tau e^{-2\pi i a\tau+2\pi
 i\frac{b_n}{\tau}} = (-i)\int_{-\infty}^{0(+)} \frac{d\tau}{\tau^2} e^{-2\pi  b_n \tau-2\pi
 \frac{a}{\tau}} \, ,
  \ee%
with
$$a=\Delta-\frac{c_L}{24}\, ,\qquad \bar a=\bar\Delta-\frac{c_R}{24}\, , $$
$$ b_n=-\Delta_n+\frac{c_L}{24}\, ,\qquad \bar b_n=-\bar\Delta_n+\frac{c_R}{24}\, .$$
In our conventions, $\Delta_n>\Delta_m$ if $n>m$ (and similarly for
$\bar\Delta$'s). In particular, the ground state energy $\Delta_0$
is the smallest eigenvalue of $L_0$. We stress that having a
non-singular CFT with a mass gap and discrete spectrum, besides in
\eqref{rho-def}, has also been used in arriving at
\eqref{rho-integral}.

One may perform the integral \eqref{a-sign} using contour integrals
and the result will be independent of the details of the contour by
virtue of Cauchy's theorem \cite{Farey-tail-1}. It can be shown that
for $a\leq 0$ ($\bar a\leq 0$) the integral \eqref{a-sign} is zero
 and thus we restrict ourselves to positive $a$ ($\bar a$) only.
Depending on the sign of $b_n$, the integral $I(a,b_n)$ is either of
the form of a Bessel function of the first kind $J_{n}(z)$  or its
modified version $I_{n}(z)$, for negative or for positive $b_n$,
respectively. Thus, using $(8.412.2)$ of \cite{G-R}, we recast
\eqref{a-sign} as
 \be \label{maro33}
I(a,b_n)=\left\{\begin{array}{ll} \frac{-2\pi\ b_n}{\sqrt{a \ b_n}}\
I_1(4\pi\sqrt{a \ b_n})\, , &\qquad b_n>0\, ,\cr \ \ \ \ \ & \ \ \ \
\cr \frac{2\pi |b_n|}{\sqrt{a |b_n|}}\ J_1(4\pi\sqrt{a |b_n|})\, ,
&\qquad b_n<0\, .
\end{array}
\right.
\ee%

 Plugging the integrals \eqref{maro33} into \eqref{rho-integral}, one obtains the recursion formula for $\rho$  %
\be\label{recursion}%
\begin{split}%
\rho(\Delta,\bar\Delta)&=(2\pi)^2\sum_{\Delta_n<\frac{c_L}{24},\bar\Delta_n<\frac{c_R}{24}}
\rho (\Delta_n,\bar\Delta_n) \frac{|\frac{c_L}{24}-\Delta_n|}{u_n}
 I_1(4\pi u_n)\cdot \frac{|\frac{c_R}{24}-\bar\Delta_n|}{v_n} I_1(4\pi v_n)\\
 &-(2\pi)^2\sum_{\Delta_n>\frac{c_L}{24},\bar\Delta_n<\frac{c_R}{24}}
\rho (\Delta_n,\bar\Delta_n) \frac{|\frac{c_L}{24}-\Delta_n|}{u_n}
 J_1(4\pi u_n)\cdot \frac{|\frac{c_R}{24}-\bar\Delta_n|}{v_n} I_1(4\pi v_n)\\
 &-(2\pi)^2\sum_{\Delta_n<\frac{c_L}{24},\bar\Delta_n>\frac{c_R}{24}}
\rho (\Delta_n,\bar\Delta_n) \frac{|\frac{c_L}{24}-\Delta_n|}{u_n}
 I_1(4\pi u_n)\cdot \frac{|\frac{c_R}{24}-\bar\Delta_n|}{v_n} J_1(4\pi v_n)\\
&+(2\pi)^2\sum_{\Delta_n>\frac{c_L}{24},\bar\Delta_n>\frac{c_R}{24}}
\rho (\Delta_n,\bar\Delta_n) \frac{|\frac{c_L}{24}-\Delta_n|}{u_n}
 J_1(4\pi u_n)\cdot \frac{|\frac{c_R}{24}-\bar\Delta_n|}{v_n} J_1(4\pi v_n) \, ,
\end{split}
\ee%
where%
\be
u_n=\sqrt{|\frac{c_L}{24}-\Delta_n|(\Delta-\frac{c_L}{24})}\, ,\qquad
v_n=\sqrt{|\frac{c_R}{24}-\bar\Delta_n|(\bar\Delta-\frac{c_R}{24})} \, .
\ee

Equation \eqref{recursion}, one of our main results, provides a recursive exact formula for the density of states and in this respect it may be viewed as analog of
the Rademacher expansion for a generic unitary, modular invariant and non-singular $2d$ CFT which is not necessarily holomorphic or holomorphically factorizable. 

To obtain the expression for the density of states we should solve
the above recursive equation for $\rho$. Recalling the behavior of
$J_{1}(z)$ and  $I_{1}(z)$ for
large arguments, %
\be\label{large-z}%
 J_1(z)\sim \sqrt{\frac{2}{\pi
z}} \sin \left(\frac\pi 4- z\right)\, , \qquad I_1(z)\sim \frac{1}{\sqrt{2\pi z}} e^z \, ,\qquad
z\gg 1\, , %
\ee %
one can show that, despite the fact that the last three lines of
\eqref{recursion}  contain $J_{1}(z)$ and are sums over infinitely
many states with (presumably) exponentially growing weights
$\rho(\Delta_n,\bar\Delta_{\bar n})$, as a result of the oscillatory
behavior of $J_1(z)$, they are nevertheless exponentially suppressed
compared to the first line for large $u_n$ and $v_n$. The details of
this analysis have been gathered in Appendix \ref{J-suppressed}.
Therefore, up to exponentially suppressed contributions, the
right-hand side of
\eqref{recursion} is given by its first line, namely %
\be\label{recursion-approx-1}%
\rho(\Delta,\bar\Delta)=(2\pi)^2\sum_{\Delta_n<\frac{c_L}{24},\bar\Delta_n<\frac{c_R}{24}}
\rho (\Delta_n,\bar\Delta_n) \frac{|\frac{c_L}{24}-\Delta_n|}{u_n}
I_1(4\pi u_n)\cdot \frac{|\frac{c_R}{24}-\bar\Delta_n|}{v_n}
I_1(4\pi v_n)\,.
\ee%

Due to the exponential behavior of $I_{n}(z)$, we can safely deduce
that in the {\em saddle point approximation} $\Delta\gg c,\
\bar\Delta\gg \bar c$, only the $n, \bar n=0$ terms, corresponding
to the maximum value of $u_n$ and $v_n$, dominate in the sum
\eqref{recursion-approx-1}, again up to exponentially suppressed
contributions.\footnote{Recalling \eqref{large-z}, the ratio of
$n$'th and zeroth terms in the sum \eqref{recursion-approx-1} for
large $u_n$ or $v_n$ is proportional to $e^{-4\pi (u_0-u_n)}
e^{-4\pi (v_0-v_n)}$. In the saddle point approximation $u_0-u_n\gg
1$ and $v_0-v_n\gg 1$.} Hence in the saddle point approximation we
can write the density of states $\rho$ as
\be\label{rho-exp-supp}%
\rho(\Delta,\bar\Delta)\simeq\left(\frac{\pi^2}{3}\right)^2\rho_0\,
{\tilde c}_{L}\frac{I_1( S_L^0)}{S_L^0}\cdot {\tilde
c}_{R}\frac{I_1( S_R^0)}{S_R^0}\,,
\ee%
where $\rho_0=\rho (\Delta_0,\bar\Delta_0)$ is the degeneracy of the
ground state which is taken to be equal to one, and
 \be\label{SL-SR}
 S_L^0=2\pi\sqrt{\frac{{\tilde{c}}_{L}}{6}\left(\Delta-\frac{c_L}{24}\right)}\, ,\qquad
 S_R^0=2\pi\sqrt{\frac{{\tilde c}_{R}}{6}\left(\bar\Delta-\frac{c_R}{24}\right)}\, ,
 \ee
 with \footnote{
It is  usually assumed that the ground state energy is zero,
\emph{i.e.} $\Delta_0=\bar{\Delta}_0=0$, for which case $\tilde
c_L=c_L,\ \tilde c_R=c_R$. However, there are interesting and
important examples with  non-zero $\Delta_0$. One famous example
discussed  in \cite{Carlip, Kutasov-Seiberg} is the Liouville
theory, for which $\Delta_0$ is such that $\tilde c=1$. Another case
with non-zero $\Delta_0$, occurs in the context of AdS$_3$ hairy
black holes \cite{Correa:2010hf}. The linear dilaton CFT (see
chapter 2  of \cite{Polchinski-book}) provides another example of
such theories. Consider $D$-dimensional bosonic string on the linear
dilaton $\Phi=V_\mu X^\mu$ background. The central charge of the
corresponding worldsheet CFT is $c=D+6V^\mu V_\mu$
\cite{Polchinski-book}. A straightforward and explicit calculation
shows that the spectrum of this theory is the same as the one of a
free bosonic string at $V_\mu=0$, except for the fact that the
zero-point energy is shifted by $\Delta_0=V^\mu V_\mu/4$.
Nonetheless, as is also stressed in footnote 1 of section 7 of
\cite{Polchinski-book}, the density of states of this theory is the
same as the one of a string theory at $V_\mu=0$. This may be
understood through our equations noting that in this case $\tilde
c=D+6 V^\mu V_\mu-24\Delta_0=D$ and
$\Delta-\frac{c}{24}=\Delta_{V=0}+\frac{V^\mu
V_\mu}{4}-\frac{D+6V^\mu V_\mu}{24}=\Delta_{V=0}-\frac{D}{24}$,
which
is $V_\mu$ independent. Linear dilaton theory is an example of a non-singular CFT which is not necessarily rational. We would like to thank the anonymous referee for a comment on this point.  }%
\be%
{\tilde c}_{L}=c_L-24\Delta_0\, ,\qquad \tilde
c_{R}=c_R-24\bar\Delta_0 \, .%
\ee

We emphasize that \eqref{rho-exp-supp} captures all polynomial
corrections to the Cardy formula, including its logarithmic
corrections. It also establishes what we set to prove
in this section, namely that any $2d$ CFT is holomorphically
factorizable in the saddle point approximation up to exponentially
suppressed contributions, {\em i.e.} the density of states is the
product of the density of states of the left and right sectors,
respectively. Moreover, as explicitly manifest in
\eqref{rho-exp-supp}, $\rho(\Delta,\bar\Delta)$ depends on the
combination of $S_L^0$ and $S_R^0$ or, rigorously, it is of the form
$\rho(\Delta,\bar\Delta)=\tilde c_L\cdot \tilde c_R\ h(S_L^0)\cdot
h(S_R^0)$. The prefactor $(\tilde c_L\cdot \tilde c_R)$ may be
understood recalling that $\rho$ is the density of states whilst the
number of states $d{\cal N}=\rho(\Delta,\bar\Delta) d\Delta
d\bar\Delta$ in the ranges $(\Delta,\Delta+d\Delta)$ and
$(\bar\Delta,\bar\Delta+d\bar\Delta)$ only depends on combinations
of $S_L^0$ and $S_R^0$ .

We conclude this section by computing the density of states at
energy $E$. For given states labeled by $\Delta_n$ and $\bar\Delta_{\bar
n}$ $(\Delta_n,\bar\Delta_{\bar n}\geq 0$), the energy $E$ and
angular momentum $J$ are defined as
\be\label{E-Jvs-Delta}%
\Delta-\frac{c_L}{24}=\frac12(E+J)\, ,\quad \bar\Delta-\frac{c_R}{24}=\frac12(E-J)\, ,
\ee%
where
$$-(E+\frac{c_L}{12})\leq J\leq E+\frac{c_R}{12}\, ,
\ \ E+\frac{c_L+c_R}{24}\geq 0\, .
$$
Next, we note that%
\be\label{boh}
\rho(\Delta,\bar\Delta)=\frac{1}{2}\frac{d^2{\cal N}}{d\Delta d\bar\Delta}=\frac{d^2 {\cal N}}{dE dJ}\, ,
\ee%
where ${\cal N}$ is the total number of states. One may then compute
$d{\cal N}/dE$ by integrating \eqref{rho-exp-supp} over $J$. We are
only interested in high energy contributions, $E\gg \frac{c_L}{24},\
\frac{c_R}{24}$, for which $-E\leq J\leq E$ and hence up to
exponentially suppressed contributions we have
\be\label{J-integrated}%
\begin{split}
\rho(E)\equiv \frac{d{\cal N}}{dE}&=\frac12\int_{-E}^E dJ\
\rho(\Delta,\bar\Delta)=\frac{\pi^2}{3}\sqrt{\tilde c_{L}\cdot
\tilde c_{R}}\int_0^{\frac{\pi}{2}} d\theta\
I_1(2\pi\sqrt{\frac{\tilde
c_{L}E}{6}}\sin\theta)I_1(2\pi\sqrt{\frac{\tilde
c_{R}E}{6}}\cos\theta)\ \cr &=\frac{\pi^2}{3}
{c_{\textrm{tot}}}\cdot \frac{I_1(S_{\textrm
{Cardy}})}{S_{\textrm {Cardy}}}+\mathrm{exponentially\ suppressed\ contributions}\, ,
\end{split}
\ee%
where%
\be\label{Cardy-S}
S_{\textrm{Cardy}}=2\pi\sqrt{\frac{c_{\textrm{tot}}E}{6} }\, ,\qquad c_{\textrm{tot}}=\tc_L+\tc_R\, .
\ee%
The details of the computation of the  integral \eqref{J-integrated} are given in Appendix \ref{Bessel-Integral}. Thus,  it is evident from \eqref{J-integrated}  that $\rho(E)$, which is the extension of the standard Cardy formula to all orders in perturbation\footnote{It is an all-orders result in the sense that we are dropping terms of the type $\exp{(-\alpha \sqrt{E})}$ and $\exp(-\beta/E)$, with $\alpha,\beta>0$.} in power of $1/E$, has the same functional form as the generic density of states \eqref{rho-exp-supp}. In particular, we note that its dependence on $E$ only appears through the $c_{\textrm{tot}}E$ combination.

\section{Cardy Formula to All Orders: Extension of the Saddle Point Analysis}\label{Simeon-method-section}

In the previous section, under natural but at the same time strong
assumptions of modular invariance, unitarity and non-singularity, we
computed the density of states for any $2d$ CFT to all orders in
perturbation theory around the saddle point, up to exponentially
suppressed contributions. In what follows, we employ the saddle
point method as used by Carlip \cite{Carlip} to go beyond his
first-order analysis and reproduce our result of the previous
section. In subsection \ref{SH},  using the modular
$S$-transformation $\beta \to 4\pi^2 / \beta$, we first {generate}
Carlip's result  \cite{Carlip} on the logarithmic correction  to the
Cardy formula. Then, in subsection \ref{beyond-log-subsection}, we
extend Carlip's analysis to extract the leading-order  correction as
well as all the subleading contributions to the Cardy formula in the
asymptotic regime where $\beta E$ is large. Finally, in subsection
\ref{E-J-section}, we consider the full partition function,
including  both energy $E$ and angular momentum $J$. Under the same
assumptions as before, we retrieve  \eqref{rho-exp-supp}. Although
the calculation is somewhat more involved, we find it helpful for
two reasons: It shows more explicitly the notion of ``up to
exponentially suppressed contributions around the saddle point'' and
it provides us with a cross-check of the result derived in section
\ref{Carlip-Extension-section}. We remark that  in this section $E$
is the energy of a state of a $2d$ CFT on a cylinder whereas
$\Delta$ and $\bar\Delta$ of the previous section are the conformal
weights of states of the CFT on $\mathbb{R}^2$ and hence they differ
by $\frac{c_L+c_R}{24}$.

\subsection{Logarithmic Corrections\footnote{The analysis in this subsection was
communicated to the authors by Simeon Hellerman  through earlier
email exchanges.}} \label{SH}

We begin by writing the partition function in the usual fashion as
 \be \label{partfunc}
  Z(\beta) = \sum_n \exp(- \beta E_n) = \int dE~ \rho(E)~\exp(-\beta E)\, ,
  \ee
where the spectral density $\rho(E)$ is a sum of delta functions
with positive integer coefficients. In spite of the fact that $\rho(E)$ is a
discrete sum of delta functions, in the asymptotic regime where $E$
is very large, the levels are very
 dense and hence is physically meaningful to approximate $\rho(E)$ with a smooth
 distribution of eigenvalues such that
\be \label{rho1st}
\rho(E) = \exp\left(\sigma(E)\right) \, .
\ee

We shall further assume that the error we may incur due to this approximation is significantly negligible compared to the error coming from truncating our expansion in corrections to the saddle point approximation at any order of our choice. In other words, we are suggesting to consider \eqref{rho1st}  as an exact equality for some smooth function $\sigma(E)$. This last remark allows us to rewrite \eqref{partfunc} as (recall the second equality in \eqref{partition-function-tau})
\be \label{partfunc2}
Z(\beta) = \int~dE~\exp(\sigma(E) - \beta E)\, ,
 \ee
and consider \eqref{partfunc2} as an equality at any order in the large-$\beta$ expansion.

In order to proceed, we draw on modular invariance to match the low-temperature expansion of $Z(\beta)$
 \be
 Z(\beta) = \exp(- \beta E_0) + a_1\exp(- \beta E_1) + {\mathcal O}(\exp(- \beta E_2))\, ,
 \ee
 as $\beta \to \infty$ to its high-temperature expansion
 \be
 Z(\beta) =
 \exp(- 4\pi^2 E_0 / \beta) +a_1 \exp( - 4\pi^2 E_1 / \beta) +{\mathcal O}(\exp(- 4\pi^2 E_2 / \beta))\, ,
 \ee
 as $\beta \to 0^+ $. Furthermore, let us make an \it ansatz \rm for the form of $\sigma(E)$. In reality,
 since we want to reverse-engineer $\sigma(E)$ to obtain the correct saddle point expansion, we ought to make an \it ansatz \rm for $\sigma^\prime(E)$ rather than for $\sigma(E)$
 \be \label{ansatz1}
  \sigma^\prime (E) = E^{q-1}\cdot
 \left(b_0 +  b_1 E^{-p}+   b_2 E^{-2p} \cdots\right)\, ,
 \ee
 where $q$ is the leading power of $E$ in $\sigma(E)$ and $p$ is some power greater than zero. Equation \eqref{ansatz1} gives {either} the following high energy expansion for $\sigma(E)$
 \be
 \sigma(E) = \ln(K)+  \sum_{m\geq 0} {{b_m}\over{q - m p }}~
 E^{q - m p}\, ,
 \ee
 if $q$ is not a positive integer multiple of
$p$, or
\be\label{sigma-expansion}%
 \sigma(E) = \ln(K) +  b_{q/p}~  \ln(E) + \sum_{\substack{m\geq 0 \\ m \neq q/p}} {{b_m}\over{q - m p }}~ E^{q - m
p }\, ,
 \ee
 if $q$ \it is \rm a positive integer multiple of $p$. In the last two equations,  $K$ is meant to be taken as a constant independent of
$E$.

With the stage now set, we are ready to attempt the evaluation of
the integral \eqref{partfunc2} in the high-temperature expansion
$\beta\to 0^+$ in terms of the unknown numbers $p,q$ and $b_m$. The
saddle point for the integral occurs   at the energy $E_*$
satisfying %
\be\label{saddle2}%
\sigma^\prime (E_*) = \beta \, .%
\ee%
 From \eqref{ansatz1} and \eqref{saddle2} we then obtain the
leading-order saddle point equation%
\be \label{saddle1}%
 b_0 E_*^{q - 1} = \beta\, , %
 \ee%
which yields the following leading-order solution\footnote{Since
both the right-hand side  of \eqref{saddle1} and $E_*$ are positive,
so  must $b_0$.}%
\be \label{saddle3}%
 E_* = (\beta / b_0)^{- {1\over{1 - q}}}\, . %
\ee%

From \eqref{saddle3}, we learn that $q$ must be less than one since $E_*$ must increase as $\beta\to 0^+$. Since
 at leading order the value of $\sigma(E_*) - \beta E_*$ ought to be equal to $\ln(Z(\beta))$, we finally have
 \be \label{saddle4}
{{1 - q}\over q} b_0^{{1\over{1 - q}}}   \beta^{- {q\over{1 -
q}}} +{\mathcal O}\left(\beta^{{p\over{1 - q}} - 1}\right) =  -
\frac{4\pi^2 E_0}{\beta}+ {\mathcal O}\left(e^{- \frac{4\pi^2 (E_1 -
E_0)}{\beta}}\right)\, .
 \ee

We draw the reader's attention to the absence in the right-hand side
of \eqref{saddle4} of power-law (or logarithmic) contributions in
$\beta$. This fact entails that the terms subleading  to
$\beta^{-1}$ at small $\beta$ vanish to all orders in perturbation
theory in fluctuations around the saddle point. We shall  use this
observation to bootstrap the higher-order terms in $\sigma (E)$.

As it may readily be seen, the limit $\beta \to 0^+$ in
\eqref{saddle4} predicts that $q =  {1\over 2}$ and
\be
b_0 = 2\pi
\sqrt{- E_0}=2\pi\sqrt{|E_0|}\, ,
\ee
as $E_0 = - {{c_L + c_R}\over{24}}$
in any unitary CFT. Thus, the leading high energy expression for
$\sigma(E)$ is
 \be \label{highsigma}
  \sigma(E) =
 4\pi\sqrt{E|E_0|} = 4\pi \sqrt{{{(c_L  +c_R) E}\over{24}}} = 2\pi
\sqrt{{{(c_L +  c_R) E}\over{6}}}   +{\mathcal O}(E^{{1\over 2} -p})\, ,
\ee
which gives the usual Cardy formula for $\rho(E)$
 \be \label{cardys1}
   \rho(E)
\simeq \exp\left\{  2\pi \sqrt{{{(c_L  + c_R) E}\over{6}}}\right\}\, .
 \ee
In fact, substituting in \eqref{cardys1} the leading-order saddle
point value, we do indeed find
 \be
   \sigma(E_*) - \beta E_* = - {{4\pi^2 E_0}\over \beta}\, ,
  \ee
 which shows we have made no mistakes at this order.

We now move on and compute the first-order correction.  We have to
be cautious about two possible sources of error in \eqref{saddle4}.
First, an error might come from including either the term
${{b_1}\over{q - p}} E^{q-p} = {{b_1}\over{{1\over 2} - p}}
E^{{1\over 2} - p}$   in the expansion for $\sigma(E)$, if
$p\neq{1\over 2}$, or the term $b_1 \ln(E)$, if $p={1\over 2}$. The
second possibly relevant source of error is the inclusion of  the
first correction to the saddle point approximation in the expression
\eqref{partfunc2} for the partition function $Z(\beta)$. Let us look
at the latter contribution first, which is nothing other than a
Gaussian integral over fluctuations of $E$ about $E_*$. Defining
$\epsilon \equiv E -E_*$, and expanding around the saddle point
$\epsilon = 0$, we find
 \be
  \sigma_{\rm leading}(E) - \beta E = {{4\pi^2 |E_0|}\over
\beta}  - {{\beta^3 }\over{16 \pi^2 |E_0|}} \epsilon^2 +{\mathcal
O}\left (  {{\beta^5}\over{|E_0|^2}}  \epsilon^3 \right )\, ,
 \ee
  where
  \be
 \sigma_{\rm leading}(E) \equiv 2~b_0 E^{1/2} = 4\pi \sqrt{|E_0| E}\, ,
 \ee
 is the leading term in the logarithm of the density of states.   Integrating over energies yields
 \bea
   \int ~ dE ~ \exp\left(\sigma_{\rm leading}(E) - \beta E\right) & = &  \exp \left ( {{4\pi^2
 |E_0|}\over \beta} \right ) \cdot \int d\epsilon   ~\exp\left \{ -
 {{\beta^3 }\over{16 \pi^2 |E_0|}} \epsilon^2  +{\mathcal O}\left (
 {{\beta^5}\over{|E_0|^2}} \epsilon^3 \right ) \right \}{} \nonumber \\
 &=&  \exp\left ( {{4\pi^2 |E_0|}\over \beta} \right ) \cdot
 \left (   4\pi^{3/2} |E_0|^{1/2} \beta^{-3/2} \right )\cdot \left (
 1   +{\mathcal O}(\beta / \epsilon) \right )\! .
 \eea
Hence,
 \be
 F(\beta) =F_{\rm Cardy}+   F_{\rm leading-fluctuation}+  F_{\rm
leading-correction-to-\sigma}  + {\mathcal O} (\beta/\epsilon)\, ,
 \ee
where $F(\beta) \equiv \ln Z(\beta)$ and
 \bea
 F_{\rm Cardy} &\equiv& {{4\pi^2 |E_0|}\over \beta}\, ,\\
 F_{\rm leading-fluctuation} &\equiv& \ln\left(4\pi^{3/2} |E_0|^{1/2} \beta^{-3/2}   \right )\, ,
 \eea
 and $F_{\rm leading-correction-to-\sigma}$ is the leading correction to  $F(\beta)$   due to the inclusion of the $b_1$ term in $\sigma(E)$. In principle, we also ought to have included the  contribution to $F(\beta)$ coming from a shift of $E_*$ as a function of $\beta$. However, we shall see that this contribution vanishes at leading order. By modular invariance, $F(\beta)$ must equal $F_{\rm Cardy} ý+{\mathcal O}\left ( \exp(- 4\pi^2 (E_1 - E_0) / \beta) \right )$ as $\beta \to 0^{+}$. From this last observation, we learn that the leading  term in $F_{\rm leading-correction-to-\sigma}$ must precisely cancel $F_{\rm leading-fluctuation}$ in the limit $\beta\to 0^{+}$, since  the latter goes as $\ln\beta$, which is much larger  than the correction of size ${\mathcal O}\left ( \exp(- 4\pi^2 (E_1 - E_0) / \beta) \right )$ appearing in the right-hand side of the equation for modular invariance \eqref{saddle4}.

Having established that first-order corrections to the saddle point
approximation  vanish and thus do not act as possible sources of
error, we turn to computing the leading correction to the value of
$E_*$ when the $b_1$ term is included. The saddle point  equation
\eqref{saddle1} is then  modified to%
\be \label{saddle5}%
 E_*^{-1/2}+{{b_1}\over{b_0}} E_*^{-1/2 - p} = {{\beta}\over{b_0}}\, ,
 \ee
where $b_0 = 2\pi \sqrt{|E_0|}$. In order to find the first-order shift in $E_*$, we  separate $E_*$ into a zeroth-order   piece and a correction piece as
 \be
 E_* = E_*^{(0)}  + E_*^{(1)}\, ,
 \ee
and expand the $b_1$-corrected saddle point equation, treating
$E_*^{(1)}$ as a small quantity to be included only at first
order.\footnote{This treatment will be justified {\it a posteriori}
at large $\beta$.}

Let us rewrite the saddle point equation \eqref{saddle5} in a more
appealing form as %
\be
  E_*\cdot \left(1   +{{b_1}\over{b_0}} E_*^{ - p}\right)^{-2} = {{b_0^2}\over{\beta^2}}\, ,
\ee
more suitable to be expanded as
 \be \label{alg1}
   \left(E_*^{(0)}+E_*^{(1)}\right)\left(1 - 2 {{b_1}\over{b_0}} \left(E_*^{(0)}\right)^{ - p}\right) = {{b_0^2}\over{\beta^2}} \, .
 \ee%
With a little algebra on \eqref{alg1}, we find that the solution for
$E_*^{(1)}$ is%
 \be
 E_*^{(1)} =   2 {{b_1}\over{b_0}}\left(E_*^{(0)}\right)^{1-p} + {\mathcal O}\left( \left(E_*^{(0)}\right)^{1-2p}\right) \, .
 \ee

At this point, we  investigate the two possible values $p$ can take
which are either $p\neq 1/2$ or $p = 1/2$. There are two corrections to
the saddle point value of the exponent $\sigma(E_*) - \beta E_*$.
The first correction%
\be%
 {{b_1}\over{{1\over 2} - p}} \left(E_*^{(0)}\right)^{{1\over 2}
- p}  +{\mathcal O} \left(\left(E_*^{(0)}\right)^{{1\over 2} - 2 p}
\right )\, ,%
 \ee%
originates from the inclusion of ${{b_1}\over{{1\over 2} - p}}
E_*^{{1\over 2} - p}$ in the expression for $\sigma(E_*)$, whilst
the second correction%
\be%
    E_*^{(1)} \cdot \left
[ \sigma^\prime\left(E_*^{(0)}\right) - \beta+{\cal O}\left(
\sigma^{\prime\prime}\left(E_*^{(0)}\right) \right) \right] \, ,%
\ee%
is a consequence of the shift in value of $E_*$.  However, by virtue
of the leading-order saddle point equation for $E_*^{(0)}$, the
second correction vanishes to the order of interest so that the
total leading contribution to $F(\beta)$ from the $b_1$ term is
 \bea
 F_{\rm leading-correction-to-\sigma} &=&  {{b_1}\over{{1\over 2} -
 p}}  \left(E_*^{(0)}\right)^{{1\over 2} - p} + {\mathcal O}\left(\left(E_*^{(0)}\right)^{{1\over 2} - 2 p} \right) \nn\\
 &=&{{b_1}\over{{1\over 2} - p}} b_0^{1 - 2p} \beta^{2p-1}  +{\mathcal O}\left (\beta^{4p-1} \right )\, ,
  \eea
 for $p\neq {1\over 2}$. Thus, we gather that the value of $p$ cannot be smaller than $1/2$, otherwise there would be a nonvanishing term of order $\beta^{2p -1}$ in $F(\beta)$, which must be absent.  Conversely, $p$ cannot be greater than $1/2$: If that were the case, the $\ln (\beta) $ term in $F(\beta )$ coming from the  fluctuation integral could not be canceled by the $b_1$ term in $\sigma(E) $.  Hence, it follows that $p$ must be exactly equal to $1/2$, and that the  leading large-$E$ behavior of $\sigma(E)$ must be
 \be
  \sigma(E) = 4\pi\sqrt{|E_0|E}+   b_1 \ln(E) +\ln(K) +{\mathcal O}(E^{-1/2})\, .
 \ee

 The partition function is then
\begin{equation}
\label{b12}
Z(\beta)=K\int_{\Lambda_{*}} dE \, \textrm{exp}\left[-\beta \big(
\sqrt{E}-b_{0}/\beta\big)^{2}+\frac{b_{0}^{2}}{\beta}\right]E^{b_{1}}\, ,
\end{equation}
where $\Lambda_{*}$ is an IR cut-off. It is clear that $Z(\beta)$ is
independent of $\Lambda_{*}$ as we are in the high $T$ (low $\beta$)
regime: A different choice of $\Lambda_{*}$ would only affect the
lower part of the spectrum of the theory in
consideration. Introducing $x=\sqrt{E}-b_{0}/\beta$, (\ref{b12}) reduces to%
 \bea\label{b123}
 Z(\beta)&=&K\int_{-\infty}^{\infty} dx \, 2\Big( x+\frac{b_{0}}{\beta} \Big)^{2b_{1}+1}\exp\left[ -\beta
 x^{2}+\frac{b_0^2}{\beta}\right]\nonumber\\
 &=&2K\exp\left[\frac{b_0^2}{\beta}\right]\left(\frac{b_{0}}{\beta} \right)^{2b_{1}+1}\int_{-\infty}^{\infty} dx\ e^{ -\beta
 x^{2}}\left(1+{\mathcal O}(\beta)\right)\nonumber\\
 &=&2K\exp\left[\frac{b_0^2}{\beta}\right]\left(\frac{b_{0}}{\beta} \right)^{2b_{1}+1}\left(\sqrt{\frac{\pi}{\beta}}+{\mathcal
 O}(1)\right)\, .
 \eea

The first term in (\ref{b123}), namely $\textrm{exp}\Big( b_{0}^{2}/
\beta \Big)$, is just the expression of the entropy as given by the
Cardy formula. Hence, by matching (\ref{b123}) to the $\beta \to
0^{+}$ limit, we are led to solve
\begin{equation}
2K\Big( \frac{b_{0}}{\beta}\Big)^{2b_{1}}\Big(
\frac{b_{0}}{\beta}\Big)\sqrt{\frac{\pi}{\beta}} =1 \, ,
\end{equation}
 from which we  find that
 \bea
 b_1 &=& - {3\over 4}\, ,\\
  K &=& 2^{- 1/2} |E_0|^{ 1/4}\, ,
 \eea
in agreement with the values found in \cite{Carlip}.

\subsection{Beyond Logarithmic Corrections}\label{beyond-log-subsection}

Thus far in this section we have reproduced the logarithmic
correction to the Cardy formula. However, we wish to go beyond the
first-order correction and reproduce the expression
\eqref{J-integrated} which is valid to all orders in the
perturbative expansion in $1/\tau$. We do so by introducing yet a
new variable $y=\sqrt{\beta} x$ which transforms the partition
function \eqref{b12} into
 \be \label{Farr}
 Z(\beta)  = 2K\frac{e^{\frac{b_0^2}{\beta}}}{\sqrt b_0}\int_{-\Lambda_{\rm UV}}^{\Lambda_{\rm UV}}
{d}y\ e^{-y^2}P(y;\beta)\, ,
 \ee
where, recalling \eqref{sigma-expansion} with $p=q=1/2$,%
\be \label{Far1}
 P(y;\beta)=\Bigg(1+\frac{y\sqrt{\beta}}{b_0}\Bigg)^{-1/2}\prod_{m=2}^\infty
 \exp{\left\{\frac{2b_m}{1-m}\left(\frac{\beta}{b_0}\right)^{m-1}
 \left(1+\frac{y{\sqrt\beta}}{b_0}\right)^{1-m}\right\}}\, ,
 \ee
 is a polynomial which can be expanded in $\frac{y\sqrt{\beta}}{b_0}$  if $\Lambda_{\rm UV}\lesssim\frac{b_{0}}{\sqrt{\beta}}$ as $\beta\to 0^{+}$.

Because of the Gaussian weight, the energy levels  above the cut-off have a vanishing contribution to the partition  function as $\Lambda_{\rm UV}\to\infty$. Thus, we can rewrite \eqref{Farr} as
 \bea
 Z(\beta)  &=& 2K\frac{e^{\frac{b_0^2}{\beta}}}{\sqrt b_0}\int_{-\infty}^{\infty}
{d}y\ e^{-y^2}P(y;\beta)\nn\\
&=&2K\sqrt{\frac{\pi}{b_0}}e^{\frac{b_0^2}{\beta}}\left(1+\sum_{k=1}^{\infty}d_k\beta^k\right)\, ,
 \eea%
where $d_k$ is a function of $b_0$ and $b_m$ for $m=2,\cdots,k-1$.
Modular invariance, as discussed in the previous subsection, implies that%
\be\label{d-k-coeff}
 d_k=0\, ,\hspace{1cm}k\in{\mathbb N}\, .
\ee
One can then use \eqref{d-k-coeff} to  determine the coefficients $b_m$ for all $m$, leading to an all-orders result.
It can  be shown that
 \be
 b_m=\frac{c_m}{b_0^{m-1}}\, ,
 \ee
 where the constants $c_m$ are the coefficients of the asymptotic expansion of the logarithm of the Bessel function $I_1(z)$
 \bea
 I_1(z)& = &\frac{e^z}{\sqrt{2\pi
 z}}\sum_{k=0}^\infty\frac{(-1)^k}{(2z)^k}\frac{\Gamma\left(\frac{3}{2}+k\right)}{k!\Gamma\left(\frac{3}{2}-k\right)}\nn \\
 &\simeq & \exp\left[z-\ln\left(\sqrt{2\pi z} \right)-\frac{3}{8z}-\frac{3}{16z^2}-\frac{21}{128z^3}-\frac{27}{128z^4}+{\mathcal
 O}(\frac{1}{z^5})\right]\, .
 \eea

 In order to put our findings in a suggestive form and make contact with \eqref{J-integrated}, we rewrite
 \begin{equation}
 \sigma(E)=\ln(K)-\frac{3}{4}\ln(E)+\sum^{\infty}_{\substack{m\geq 2}}
  \,\frac{2b_{m}}{1-m}E^{\frac12(1-m)}\, ,
 \end{equation}
 as\footnote{We recall that $|E_0|=\frac{c_{\rm tot}}{24}$, $K=\frac{1}{2}\sqrt{\frac{b_0}{\pi}}$ and
 $b_0=2\pi\sqrt{|E_0|}$.}
 \bea
 \sigma(S_{\textrm{Cardy}})&=&S_{\textrm{Cardy}}
 +\ln \left(\frac{\pi^2}{3}{c_{\rm tot}}\right)
 -\frac{3}{2}\ln\left({S_{\rm Cardy}}\right)-\sum_{m=2}^\infty
 \frac{2^mc_m}{m-1}\left(\frac{1}{S_{\rm Cardy}}\right)^{m-1}\nn \\
 &=&\ln \left(\frac{\pi^2}{3}c_{\rm tot}\right)-\ln\left({S_{\rm Cardy}}\right)+\ln(I_1(S_{\rm Cardy}))\, ,
 \eea
or
\be\label{Simeon-method-final}
\rho(E)= e^{\sigma }=\frac{\pi^2}{3}\frac{c_{\rm tot}}{S_{\rm Cardy}}I_1(S_{\rm Cardy})\, ,
\ee
where $S_{\rm Cardy}$ and $c_{\rm {tot}}$ are defined in \eqref{Cardy-S}.
Equation \eqref{Simeon-method-final} is obviously the same as the result \eqref{J-integrated}  obtained in the previous section .

\subsection{Generic Case with $E$ and $J$}\label{E-J-section}

Ideally, we would like to extend our previous analysis and
generalize it to the case in which  $J\neq 0$ and $c_L\neq c_R$. On
general grounds, %
\be%
\sigma(\Delta,\bar\Delta)=\sum_{\substack{m=-M\\ m\neq 0}}^\infty
\frac{a_m}{m r}\Delta^{p-m r}+ \sum_{\substack{n=-N\\ n\neq
0}}^\infty \frac{{\bar a}_n}{ns} {\bar\Delta}^{q-n
s}+\alpha\ln\Delta+\beta\ln\bar\Delta+\eta(\Delta,\bar\Delta)\, , \ee
where \be \eta(\Delta,\bar\Delta)=\sum_{n \cdot m\neq
0}c_{m,n}\Delta^{mr}{\bar\Delta}^{ns}\, , %
\ee%
with $\Delta$ and
$\bar\Delta$  related to $E$ and $J$ as in \eqref{E-Jvs-Delta}.

According to the lore of modular invariance, we have\footnote{Note that in this subsection $\tau$ ($\bar\tau$) is equal to $-2\pi i\tau$ ($2\pi i\bar\tau$) of the previous section.}
\be\label{J-extended-I}
Z(\tau,\bar\tau)=\int d\Delta\
d\bar\Delta\exp\left(\sigma(\Delta,\bar\Delta)-\tau\Delta-\bar\tau\bar\Delta\right)=\exp\left\{-2\pi^2\left(\frac{E_0}{\tau}+
\frac{\bar E_0}{\bar\tau}\right)\right\}\, ,
\ee
where we have assumed $\Delta_{\rm min}=E_0/2$ and $\bar\Delta_{\rm min}=\bar E_0/2$. From our saddle point analysis, we are led to set $N=M=1$, $r=s=\frac{1}{2}$, $p=q=0$; also, $a_{-1}^{2}=-2\pi^2E_0$ and $\bar a_{-1}^{2}=-2\pi^2\bar E_0$. Equation \eqref{J-extended-I} then reads
\bea \label{maro44}
Z(\tau,\bar\tau) & = &  4\int_{x_0}^\infty dx\ \int_{\bar x_0}^\infty d\bar x \left(x+\frac{a_{-1}}{\tau}\right)^\alpha
\left(\bar x+\frac{\bar a_{-1}}{\bar \tau}\right)^\beta
e^{(-\tau x^2-\bar\tau {\bar  x}^2)}{}\\
& = & 4\int_{x_0}^\infty dx\ \int_{\bar x_0}^\infty d\bar x \left(\frac{a_{-1}}{\tau}\right)^\alpha
\left(\frac{\bar a_{-1}}{\bar \tau}\right)^\beta
\left(1+\frac{\tau x}{a_{-1}}\right)^\alpha\left(1+\frac{\bar \tau\bar x}{\bar a_{-1}}\right)^\beta
e^{(-\tau x^2-\bar\tau {\bar  x}^2)}\simeq 1 \nn\, ,
 \eea
 where $x_0=-\frac{a_{-1}}{\tau}$ and $\bar x_0=-\frac{\bar
 a_{-1}}{\bar\tau}$. Since we are working with a Gaussian integral, we can take both $x_0,\bar x_0\to-\infty$. Thus, \eqref{maro44}
 implies that $\alpha=\beta=-\frac{3}{4}$ since $\frac{1}{\tau}$ should cancel on the left hand side.

To go beyond saddle point approximation, we need to show that
$\eta(\Delta,\bar\Delta)=0$. Let us consider a term in $\eta$ like
the following%
\be%
c_{m,n}\Delta^{m/2}{\bar\Delta}^{n/2},\hspace{1cm}m\cdot n\neq0\, .
\ee %
This last term can be written as a function of $x$ and $\bar x$ as%
\be%
 c_{m,n}\left(x+\frac{a_{-1}}{\tau}\right)^{m}\left(\bar
x+\frac{\bar a_{-1}}{\bar \tau}\right)^{n}\simeq
c_{m,n}\left(\frac{a_{-1}}{\tau}\right)^{m} \left(\frac{\bar
a_{-1}}{\bar \tau}\right)^{n}P\left(\frac{\tau x}{a_{-1}},\frac{\bar
\tau\bar x}{\bar a_{-1}}\right)\, , \hspace{1cm}m\cdot n\neq0 \, ,
\ee %
where $P$ is a polynomial and $P(0,0)=1$. After taking the expansion%
\be%
\exp\left\{c_{m,n}\left(\frac{a_{-1}}{\tau}\right)^{m}
\left(\frac{\bar a_{-1}}{\bar \tau}\right)^{n}\right\}\simeq
1+c_{m,n}\left(\frac{a_{-1}}{\tau}\right)^{m} \left(\frac{\bar
a_{-1}}{\bar \tau}\right)^{n}\ , \ee we deduce from
\eqref{J-extended-I}  that \be c_{m,n}=0\, , %
\ee%
 which implies that
the theory is holomorphically factorized around the saddle point.

One can then readily continue the analysis along the lines worked out in subsection \ref{beyond-log-subsection} and reproduce \eqref{rho-exp-supp}. Since the computation is basically the same as that shown in the previous subsection we do not repeat it here.

\section{Canonical vs. Microcanonical
Entropy}\label{micro-vs-canonic-section}

In the previous two sections, starting from the canonical partition
function we derived the expression for the microcanonical density of
states through a Laplace transform. Then, employing modular
invariance of the partition function we fixed the form of the
microcanonical density of states, up to exponentially suppressed
contributions in the saddle point approximation. By means of the
usual thermodynamical equations, one can then read the
\emph{microcanonical entropy} $S_{\textrm{m.c.}}$ upon taking the
logarithm of the density of states. Let us first consider the pure
imaginary $\tau$ case, corresponding to the $L_0=\bar
L_0=E$ sector, for which the microcanonical entropy  is%
\bse%
\begin{align}%
S_{\textrm{m.c.}}&=\ln\left(\rho(E)\right)=
\ln\left(\frac{\pi^2}{3} c_{\textrm{tot}}\right)+
\ln\left(\frac{I_1(S_{\textrm{Cardy}})}{S_{\textrm{Cardy}}}\right)
\label{micro-entropy-E-1}\\ %
&= \ln\left(\frac{\pi^{3/2}}{3\sqrt{2}}
c_{\textrm{tot}}\right)+S_{\textrm{Cardy}}-\frac{3}{2}\ln
S_{\textrm{Cardy}}+{\cal
O}\left(\frac{1}{S_{\textrm{Cardy}}}\right)\, ,
\label{micro-entropy-E-2}\end{align}
\ese%
where $S_{\textrm{Cardy}}$ is given in \eqref{Cardy-S}. We note that
\eqref{micro-entropy-E-1} gives the microcanonical entropy up to
exponentially suppressed terms whilst \eqref{micro-entropy-E-2} captures
only the leading log-correction to the Cardy formula. The
logarithmic correction was also discussed in \cite{Carlip}.

Given the density of states $\rho(E)$ \eqref{J-integrated}, one may
insert it back into the expression for the partition function $Z$%
\be\label{Z-canonic}%
Z(\beta)=\int dE \ \rho(E)\ \exp(-\beta E)\ , %
\ee%
and compute the \emph{canonical entropy} $S_{\textrm{c}}$ %
\be\label{canonic-entropy}%
S_{\textrm{c}}=\ln Z-\beta\frac{\partial\ln
Z}{\partial\beta}\, .%
\ee%
The integral \eqref{Z-canonic} can be performed using formula
(6.620.4) of \cite{G-R} to obtain%
\be
Z(\beta)=\frac12\left(e^{\frac{\pi^2}{6}c_{\textrm{tot}}T}-1\right)\
,
\ee%
where $T=\beta^{-1}$ is the temperature. One then arrives at\footnote{One may compute the microcanonical temperature
$T_{\textrm{m.c.}}$ given the microcanonical
entropy:  $T_{\textrm{m.c.}}^{-1}=\frac{\partial
S_{\textrm{m.c.}}}{\partial E}$. Using \eqref{J-integrated}
we obtain
$$
\frac{\pi^2}{3}c_{\textrm{tot}}T_{\textrm{m.c.}}=
\frac{S_{\textrm{Cardy}}I_1(S_{\textrm{Cardy}})}{I_2(S_{\textrm{Cardy}})}\
,
$$
where $I_2(x)$ is the modified Bessel function and $S_{\textrm{Cardy}}$
is given in \eqref{Cardy-S}.
}%
\be\label{S-canonic-Cardy}%
S_{\textrm{c}}=\ln 2+\frac{\pi^2}{3}c_{\textrm{tot}}T+
{\cal O}(e^{-\frac{\pi^2}{3}c_{\textrm{tot}}T})\, .
\ee%
That is, the canonical entropy, in the saddle point approximation
and up to exponentially suppressed contributions, is completely
given by the Cardy formula and in particular there are no logarithm
or other polynomially suppressed terms, which appeared in the
microcanonical entropy \eqref{micro-entropy-E-1}. In other words,
the functional dependence of the microcanonical density of states
\eqref{J-integrated}, namely $\frac{1}{S}I_1(S)$, is such that the
Cardy formula is the exact expression for the canonical entropy, up
to exponentially suppressed contributions.

The above argument can be readily generalized to the case with
non-zero $J$, for generic $\Delta$ and $\bar\Delta$. As we
discussed, the microcanonical density of states \eqref{rho-exp-supp} is the product of density of states in the left and
right sectors. Inserting \eqref{rho-exp-supp} into
\eqref{partition-function-tau}  the canonical
partition function takes the form%
\be%
Z(\tau,\bar\tau)=Z_L(\tau) \cdot Z_R(\bar\tau)\, ,%
\ee%
where%
\be%
Z_L(\tau)=\frac12\left(e^{\frac{\pi i}{12}\frac{\tilde
c_L}{\tau}}-1\right),\qquad Z_R(\bar\tau)=\frac12\left(e^{-\frac{\pi
i}{12}\frac{\tilde c_R}{\bar\tau}}-1\right)\, . %
\ee%
As the details of the computations exactly parallel those of the
previous case, we do not repeat them again. The canonical entropy is then given by%
\be\label{S-canonic-E-J}%
S_{\textrm{c}}=S_{\textrm{c}}^L+S_{\textrm{c}}^R\, ,%
\ee%
where%
\be%
S_{\textrm{c}}^L=\ln Z_L-\tau\frac{\partial\ln
Z_L}{\partial\tau},\qquad S_{\textrm{c}}^R=\ln
Z_R-\bar\tau\frac{\partial\ln Z_R}{\partial\bar\tau}\, . %
\ee%
In terms of left and right temperatures $T_L=-\frac{1}{2\pi i \tau}$
and $T_R=\frac{1}{2\pi i\bar\tau}$, the canonical entropy
\eqref{S-canonic-E-J} is obtained to be\footnote{Recall that for a generic $T_L$ and $T_R$, the partition function
\eqref{partition-function-tau} may be written as
$\Tr \left(e^{-\beta_L L_0}\ e^{-\beta_R \bar L_0}\right)$ or equivalently
$\Tr \left(e^{-\beta H-\beta\mu J}\right)$, where $\beta$ is the inverse of canonical temperature,
$\mu$ is the chemical potential for the angular momentum $J$, $L_0+\bar L_0=2H$, $L_0-\bar L_0=2J$ and
$\beta_L=1/T_L,\ \beta_R=1/T_R$.
Therefore,%
$$ \frac{2}{T}=\frac{1}{T_R}+\frac{1}{T_L},\qquad
\frac{2\mu}{T}=\frac{1}{T_L}-\frac{1}{T_R}.$$}
\be%
S_{\textrm{c}}= 2\ln 2+\frac{\pi^2}{3}\left(\tilde c_L
T_L+\tilde c_R T_R\right)+ {\cal
O}(e^{-\frac{\pi^2}{3}\tilde c_LT_L},\ e^{-\frac{\pi^2}{3}\tilde c_RT_R} )\, .%
\ee%
Again,  we see that the Cardy formula  gives the exact
canonical entropy up to exponentially suppressed terms.


\section{Discussion and Outlook }\label{Discussion-Section}

In this work, we have exploited $S$-transformation of the
PSL$(2,\mathbbm{Z})$ modular group to learn more about the density
of states of a generic unitary, modular invariant and non-singular
$2d$ CFT. Interestingly, we have found that the density of states is
the product of those of the left and  right sectors and,
furthermore, it only depends on $S_0\equiv \tilde c
(\Delta-\frac{c}{24})$ with $\tilde c=c-24\Delta_0$ in each sector,
up to \emph{exponentially suppressed contributions in the saddle
point approximation}. Our main result, \eqref{rho-exp-supp}, may
have diverse interesting physical implications. We discuss some of
them below.
\begin{enumerate}
\item
Noticing that the density of states only depends on $S_0$, one can
construct a class of $2d$ CFTs with different central charges and
different spectra, but with the same $S_0$. This class of CFTs, for
which the densities of states are the same, is relevant for
``orbifolded CFTs'' or CFTs on an $\mathbbm{R}\times
S^1/\mathbbm{Z}_k$ orbifold. As an evocative and related argument,
we observe  that any Virasoro algebra at central charge $c$ has
infinitely many Virasoro subalgebras labeled by integer $k$ at
central charge $c k$. This point has also been noted in
\cite{Banados}.
    To appreciate this last comment, let us consider the algebra \eqref{Vir-algebra} and  the subset of its generators $L_{nk}$
\be
\begin{array}{l}
{\hat L}_n=\frac{1}{k}L_{nk}, \hspace{1cm}n\neq0,\\\\
\hat L_0-\frac{\hat c}{24}=\frac{1}{k}\left(L_0-\frac{c}{24}\right)\ ,
\end{array}
\label{projection} \ee where $\hat c=ck$. It is then straightforward
to see that the set of ${\hat L}_n$ also forms a Virasoro algebra at
central charge ${\hat c}$. The relation between $\hat L_n$ and $L_n$
could be understood through a (non-single valued) conformal map
$w=z^k$ on the complex plane. This map in turn, can  be viewed
as orbifolding the $z$-plane by $\mathbb{Z}_k$. According to this line of
reasoning, it can therefore be stated that $\hat c
(\hat\Delta-\frac{{\hat c}}{24})=c(\Delta-\frac{c}{24})$. This is
suggesting that two conformal field theories with central charge and
spectrum $(c,\Delta)$ and $(\hat c,\hat\Delta)$ respectively, should
have the same density of states up to exponentially suppressed
contributions.

In light of the above arguments, it might be possible to relax the requirement of having integer-valued spacing of the conformal weights $\Delta_n$ (\emph{cf.} discussions below \eqref{rho-def}) while keeping the discreteness of the spectrum: For theories with   $(kc,\Delta/k)$ one might hope to find a ``dual'' CFT with $(c,\Delta)$ (see the example below). In this ``dual'' picture the spacing of the spectrum and the mass gap $\delta M$ is then $1/k$, or in a more suggestive form $\delta M\sim 1/c$.

\item The observation in item 1. may be employed in the black
hole microstate identifications within string theory. In particular,
it may be used to relate black holes within the family of
$(D0,D6)$-branes with different number of $D6$-branes. A discussion
on the latter subject may be found in \cite{Vishnu}.  In this context
 families of $(kc,\Delta/k)$ CFTs are mapped to each other by T-duality. See also \cite{Farhang} for a related analysis.

\item
Although we mainly focused on the analysis of $2d$ CFT, one of our
main motivations to study this problem was the desire to shed new
light on the AdS$_3$/CFT$_2$ correspondence (see \cite{MAGOO} for a
review). As first pointed out by Strominger \cite{Strominger-97},
the Bekenstein-Hawking entropy of BTZ black holes \cite{BTZ} is
correctly reproduced by the Cardy formula \eqref{Cardy-density},
which is nothing but the thermodynamical entropy of a $2d$ CFT with
Brown-Henneaux central charge \cite{Brown-Henneaux} at temperature
equal to the Hawking temperature of the BTZ black hole.

As shown in details earlier (see also \cite{Carlip}), Cardy
formula should be viewed as the first term in the saddle point
expansion. The natural question is then: What do the corrections
to the Cardy formula correspond to in the gravity picture?

Of course, before employing our results to address questions in
AdS$_3$ gravity, one should make sure that the assumptions of
modular invariance and non-singularity are expected to hold for the
CFT proposed to be dual to quantum gravity in AdS$_3$. Naturally,
modular invariance is needed if we want the dual gravity to be
compatible with toroidal boundary conditions on the Euclidean
AdS$_3$, see \cite{Witten:2007kt,Maloney:2007ud}. The
non-singularity and existence of a mass gap are harder to argue for.
For example, for the $D1/D5$ system as discussed in
\cite{Seiberg-Witten}, there are some regions in the parameter space
where the dual theory is expected to have a continuous spectrum
above a gap. This brings about a pathology in the dual $2d$ CFT.
Nevertheless, in generic points of the parameter space of the
$D1/D5$ system the dual CFT is non-singular. Then, it seems
plausible to assume that the consequences of our findings may be of
significance for  quantum gravity in AdS$_3$. This is compatible
with the arguments of \cite{Witten:2007kt,{Maloney:2007ud}}.

An idea to tackle the question posed above, as motivated by stringy
$\alpha'$-corrections to supergravity (\emph{e.g.} see \cite{Sen}
for a review), is that the Bekenstein-Hawking area law receives
corrections from the higher derivative corrections to the
Einstein-Hilbert action \cite{Wald}. This idea is shown to work for
certain four- and five-dimensional black holes in higher derivative
gravity theories  \cite{Sen}. Accordingly, the corrections to the
Cardy formula  may then play the role of corrections to that action
whilst the exponentially suppressed contributions may be associated
with ``stringy or non-perturbative corrections'' which do not admit
a semiclassical description. Stated differently, the existence of
corrections to the Cardy formula implies that the effective
semiclassical description of a quantum gravity in AdS$_3$ admitting
a dual CFT description cannot  simply be AdS$_3$ Einstein gravity
without higher derivative corrections.\footnote{ It has been argued
that the log-corrections \cite{Carlip} and in general the whole
Rademacher expansion \cite{Birmingham}, may be associated with
quantum gravity effects.} However, $3d$ gravity has specific
features: It was shown by Saida and Soda \cite{SS} that any higher
derivative corrections to Einstein gravity in AdS$_3$, irrespective
of the details of such corrections, only result in a shift in the
AdS$_3$ radius and hence a shift in the Brown-Henneaux central
charge. It is obvious that our corrections to the Cardy formula
cannot be captured by only a shift in the central charge. Therefore,
this particular idea does not work and our corrections  should be
understood in a different way.

In view of the results of \cite{SS}, the answer to the above
question lies within the lines of section
\ref{micro-vs-canonic-section}. The Bekenstein-Hawking entropy of
a BTZ black hole should be viewed as entropy of a system in a
\emph{canonical ensemble}. This is due to the fact that  a
black hole as perceived by an observer at infinity is a system at a
given temperature, the Hawking temperature. This fact becomes more
apparent recalling the Wald entropy formula \cite{Wald}, in which
entropy is associated with a Noether charge while  the temperature is fixed
to be the Hawking temperature. As we showed in section
\ref{micro-vs-canonic-section}, the Cardy formula
\eqref{S-canonic-Cardy} does not receive logarithmic or polynomial
corrections in powers of $c\cdot T$, it is exact up to exponentially
suppressed contributions.

\item
An interesting and important question put forth in
\cite{Witten:2007kt}, and discussed further in \cite{Maloney:2007ud},
is whether there is a well-defined AdS$_3$ pure Einstein quantum gravity, defined by a path integral over the AdS$_3$
Einstein-Hilbert action, where the metric is the only dynamical
field with  prescribed   boundary conditions \cite{Brown-Henneaux}.
In \cite{Maloney:2007ud} a careful analysis of the path integral for pure AdS$_3$ Einstein gravity is carried out,  taking into account  contributions of all ``Brown-Henneaux states''. (Brown-Henneaux states are the boundary excitations of AdS$_3$ background which respect the Brown-Henneaux boundary conditions \cite{Brown-Henneaux}. These excitations are localized around the conformal boundary of AdS$_3$ \cite{Maloney:2007ud}.) It was then argued that this path integral  does not have the expected form of the  partition function of a ``physically sensible theory''.

Our findings may have implications on  this question/puzzle. If
there exists such an AdS$_3$ quantum theory (minimal, pure Einstein
theory, or otherwise) of gravity and it admits a modular invariant,
non-singular and unitary dual $2d$ CFT, then according to
\eqref{rho-exp-supp} this CFT is holomorphically factorizable and
one may relax demanding this last feature as a requirement. Note
also that \eqref{rho-exp-supp} is obtained from \eqref{recursion} by
discarding the exponentially suppressed contributions in the saddle
point approximation and that in the derivation of \eqref{recursion}
no assumption about holomorphic factorizability has been made.
(Equation \eqref{recursion} is an exact result which is true  
for generic $2d$ CFTs whose partition functions
are neither necessarily holomorphic nor holomorphically
factorizable.) In other words, our equation \eqref{recursion} should
be compared to equations (5.13) and  (5.4) of \cite{Maloney:2007ud},
or more precisely to the modular invariant partition function given
in equation (3.7) therein, which  gives the entropy  associated with
a BTZ black hole computed from the partition function of AdS$_3$
gravity (calculated as described above). As another outcome of this
comparison, we observe that equation (5.13) (or equation (3.7))
turns into the partition function of a ``physically sensible
theory'' (\emph{cf.} discussions of \cite{Maloney:2007ud}) if the
coefficients $C_\Delta$ are replaced with
$\rho(\Delta_n,\bar\Delta_n)$ of  our analysis. The physical meaning
of this observation and its implications will be analyzed elsewhere.

\item
In this work we mainly focused on the results ``up to exponentially
suppressed contributions in the saddle point approximation''. The
interpretation of these exponentially suppressed terms, especially
with regard to the question of AdS$_3$ quantum gravity and the
 (BTZ) black holes entropy for the cases involving holomorphic partition function, corresponding to BPS black holes, has been discussed in e.g. \cite{Farey-tail-1,exp-supp}. Given our analysis here, it is
another interesting question to study these contributions for the case of generic non-BPS black holes.

\end{enumerate}


 \section*{Acknowledgements}
It is a pleasure to acknowledge the collaboration of Simeon
Hellerman at earlier stages of this project and for inspiring the
analysis of Section \ref{Simeon-method-section} of this paper. We
would like to thank Steven Carlip, Jan de Boer and Joan Sim\'{o}n
for comments on the draft and Reza Fareghbal for making us aware of
reference \cite{SS}. MV wishes to express his heartfelt gratitude to
{\em all} members of IPM for providing a very friendly and
stimulating environment as well as support of any kind: {\it yek
donya mamnoon}.

\appendix


\section{Suppression of the Oscillatory Terms in $\rho(\Delta,\bar\Delta)$} \label{J-suppressed}

In this Appendix we  show that the last three terms in
\eqref{recursion} are indeed suppressed compared to the first term.
To this end,  we estimate these three terms and compare them
against the first term. We do so by assuming that our claim holds,
that is %
\be%
 \rho(\Delta,\bar\Delta)\simeq (2\pi)^2 {\tilde
c}_L{\tilde c}_R \frac{I_1(4\pi u_0)}{u_0}\cdot \frac{ I_1(4\pi v_0)}{v_0}\, ,
\ee where \be u_0=\sqrt{\tilde c_L(\Delta-\frac{c_L}{24})}\, , \qquad
v_0=\sqrt{\tilde c_R(\bar\Delta-\frac{c_R}{24})}\, , %
\ee %
and then replace the above expression for $\rho$ into the sums in
the last three terms in \eqref{recursion}. Noting the exponential
growth of the states by energy, the main contribution to the sums
comes from the large $\Delta$ states, a region of the spectrum in
which the sums could be approximated by integrals over $\Delta$.
The integrals one needs to compute then take the form%
\be\label{2nd-term-estimate}%
\begin{split}%
{\cal I}(a,b) &= \int_0^\infty dE\ I_1(a\sqrt{E}) J_1(b\sqrt{E})\\
&=\int_0^\infty dx\ x I_1(a x) J_1(b x)\, .
\end{split}
\ee%

To compute the above integral, we make use of $(8.447.2)$ and $(6.511.1)$
of \cite{G-R}
\be\label{Bessel-I-Bessel-J}%
\begin{split}%
I_1(az) &= \sum_{k=0}^\infty \frac{1}{k!(k+1)!}
\left(\frac{az}{2}\right)^{2k+1}\, , \\ %
\frac{1}{b} &=\int_{0}^{\infty}dx\  J_1(bx)\, .
\end{split}%
\ee%

The integral \eqref{2nd-term-estimate} then becomes
\be\label{I(a,b)-1}\begin{split}%
{\cal I}(a,b) =\sum_{k=0}^\infty\frac{1}{k!(k+1)!} \left(\frac{a}{2}\right)^{2k+1} c_{k+1}(b)\, ,
\end{split}%
\ee%
where
\be\label{c-k}%
c_{k}(b)\equiv\int_{0}^{\infty}dx\  x^{2k} J_1(bx)\, .
\ee%

If we have the expression for $c_k$ we can then compute ${\cal I}(a,b)$.

From \eqref{c-k}, one can  then show that%
\be \label{recursion-c-k}
\frac{d^2}{d^2 b}c_k(b)=-c_{k+1} -\frac{d}{db}\left(\frac{1}{b} c_k(b)\right)\, .
\ee

To obtain \eqref{recursion-c-k}, we have used the following identities for $J_{n}(z)$
\be
z\frac{d}{dz}J_1(z)=zJ_0(z)-J_1(z)\, , \quad J_1(z)=-\frac{d}{dz} J_0(z) \, .
\ee

The solutions to \eqref{recursion-c-k} are of the generic form
\be \label{recursion-c-k2}
c_k(b)=\frac{1}{b^{2k+1+2\epsilon}} d_k \, ,
\ee
where $d_k$ is  a $b$-independent parameter to be determined and $\epsilon$ is an arbitrary number. Inserting \eqref{recursion-c-k2} into
\eqref{recursion-c-k} we obtain%
\be
d_{k+1}=-4(k+\epsilon)(k+1+\epsilon) d_k\ ,\quad d_0=1 \, .
\ee

We next note that from \eqref{c-k} one can read $c_k(b)= b^{-2k-1-2\epsilon} c_k(b=1)$ if we \emph{regulate} the integral by replacing $dx$ with $d^{1+2\epsilon}x$. In other words, $\epsilon$ should be viewed as a dimensional regularization parameter which will be taken to zero at the end of the computation. Therefore, to leading order in $\epsilon$ we have
\be
d_{k+1}=(-1)^k 2^{2k} k!(k+1)! d_1\, , \ k\geq 0\, , \quad d_1=-4\epsilon \, ,
\ee
and hence%
\be\label{I(a,b)-reg}
{\cal I}(a,b)= \frac{d_1}{2b^{2(1+\epsilon)}}\sum_{k=0}^\infty (-1)^k\left(\frac{a}{b}\right)^{2k+1}=\frac{d_1}{2b^{2\epsilon}}
\frac{a}{b} \frac{1}{a^2+b^2}\, ,
\ee%
where we have assumed $b>a$ (for $a>b$ the sum is not convergent). Thus, ${\cal I}(a,b)=0$ in the $\epsilon\to 0$ limit. We would like to comment that from \eqref{2nd-term-estimate} one can show that
${\cal I}(a,b)=\frac{1}{b^2}{\cal I}(\frac{a}{b}, 1)$ and, as clearly seen  from
\eqref{I(a,b)-reg},  our regularization  respects this property.

We wish to conclude this Appendix by emphasizing that  a similar result on the vanishing of contributions from the Bessel function $J_1(z)$,
was established in Appendix B of \cite{Farey-tail-1} using a different regularization scheme.

\section{Computation of the Bessel Function Integral}\label{Bessel-Integral}

Here, we present a detailed computation of the integral \eqref{J-integrated}. Noting that $I_n(z)=i^{-n} J_n(iz)$, instead of \eqref{J-integrated} one may compute%
\be \label{J-integrated22}
J(x,y)\equiv \int_0^{\frac{\pi}{2}} d\theta J_1(x\sin\theta) J_1(y\cos\theta)\, .
\ee%

Let us sketch out the steps for doing this computation:
\begin{enumerate}

\item By means of formula $(8.535)$ of \cite{G-R}, we express the $J_{n}(z)$ in \eqref{J-integrated22} as
\be\begin{split}
J_1(x\sin\theta)&=\sin\theta\sum_{k=0}^\infty\frac{1}{k!} J_{k+1}(x) \left(\frac{x}{2}\right)^k\cos^{2k}\theta \, ,\\
J_1(y\cos\theta)&=\cos\theta\sum_{l=0}^\infty\frac{1}{l!} J_{l+1}(y) \left(\frac{y}{2}\right)^l\sin^{2l}\theta\, .
\end{split}
\ee%

\item Using formula $(3.621.5)$ of \cite{G-R}, we can perform the theta integral
\be
\int_0^{\frac{\pi}{2}} d\theta \sin^{2l+1}\theta\cos^{2k+1}\theta\ =\frac{k!l!}{2(k+l+1)!}\, ,
\ee
to derive%
\be
J(x,y)=\sum_{k,l=0}^\infty\ \frac{1}{2(k+l+1)!}\ J_{k+1}(x) J_{l+1}(y)\ \left(\frac{x}{2}\right)^k\left(\frac{y}{2}\right)^l \, .
\ee%

\item With the help of formula $(8.440)$ of \cite{G-R}
\be
J_{k+1}(x)=(-1)^{k+1}\left(\frac{x}{2}\right)^{-(k+1)}\sum_{p=k+1}^\infty\
\frac{(-1)^p}{p!\Gamma(p-k)}\left(\frac{x}{2}\right)^{2p}\, ,
\ee
we obtain
\be \label{maro}
J(x,y)=\frac{2}{xy}\ \sum_{p,q=1}^\infty\ \frac{(-1)^{p+q}}{p!q!} \left(\frac{x}{2}\right)^{2p}\left(\frac{y}{2}\right)^{2q}\cdot C_{p,q}\, ,
\ee
where%
\be\begin{split}
C_{p,q} &=\sum_{k=0}^{p-1}\sum_{l=0}^{q-1}\ \frac{(-1)^{k+l}}{(p-k-1)!(q-l-1)!}\frac{1}{(k+l+1)!}\\
&=(-1)^{p+q}\ \sum_{r=0}^{p-1}\sum_{s=0}^{q-1}\ \frac{(-1)^{r+s}}{r!s!(p+q-r-s-1)!}\, .
\end{split}
\ee%

\item One can show that%
\be C_{p,q}=\frac{1}{ (p+q-1)!}\, . \ee

\item Thus, we are able to rewrite \eqref{maro} as
\be \label{J(x,y)}%
J(x,y)=\frac{2}{x y}\sum_{p,q=0}^\infty\ \frac{(-1)^{p+q}}{(p+1)!(q+1)!(p+q+1)!} \left(\frac{x}{2}\right)^{2(p+1)}\left(\frac{y}{2}\right)^{2(q+1)}\, .
\ee%

\item Next, let us consider the expansion of $J_1({\sqrt{x^2+y^2}})$
\be\begin{split}\label{Q(x,y)}%
Q(x,y) &\equiv \sqrt{x^2+y^2}\ J_1({\sqrt{x^2+y^2}})=2\sum_{n=0}^\infty\ \frac{(-1)^n}{n! (n+1)!}\ \left(\frac{x^2+y^2}{4}\right)^{n+1}\cr
&= 2\sum_{n=0}^\infty\  \frac{(-1)^n}{n!}\ \sum_{m=0}^{n+1}\ \frac{1}{m!(n+1-m)!}\ \left(\frac{x}{2}\right)^{2m}\left(\frac{y}{2}\right)^{2(n-m+1)}\cr
&=-2\sum_{\substack{p,q=-1\\ p+q\geq -1}}^\infty\ \frac{(-1)^{p+q}}{(p+1)!(q+1)!(p+q+1)!} \left(\frac{x}{2}\right)^{2(p+1)}\left(\frac{y}{2}\right)^{2(q+1)}\, .
\end{split}
\ee%
\item To compare \eqref{Q(x,y)} with  \eqref{J(x,y)}, we decompose the above sum into three different regions,
$p,q=0,\cdots, \infty$, $p=-1,\ q=0,\cdots, \infty$ and $q=-1,\ p=0,\cdots,\infty$ such that
\be\begin{split}
Q(x,y)&=-2\sum_{p,q=0}^\infty\ \frac{(-1)^{p+q}}{(p+1)!(q+1)!(p+q+1)!} \left(\frac{x}{2}\right)^{2(p+1)}
\left(\frac{y}{2}\right)^{2(q+1)}\cr &
+2\sum_{p=0}^\infty\ \frac{(-1)^{p}}{p!(p+1)!} \left(\frac{x}{2}\right)^{2(p+1)}+2\sum_{q=0}^\infty\
 \frac{(-1)^{q}}{q!(q+1)!} \left(\frac{y}{2}\right)^{2(q+1)}\, .
\end{split}\ee%

\item Using once more $(8.440)$ of \cite{G-R}, we deduce
\be
J(x,y)=-\frac{\sqrt{x^2+y^2}}{xy} \ J_1({\sqrt{x^2+y^2}})+\frac{1}{y} J_1(x)+\frac{1}{x} J_1(y)\, .
\ee%
\end{enumerate}

We can now take $x,y$ to be imaginary-valued and finally arrive at%
\be \label{expo}
\begin{split}
I(x,y) &\equiv \int_0^{\frac{\pi}{2}} d\theta I_1(x\sin\theta) I_1(y\cos\theta)=-J(ix,iy)\cr
&= \frac{\sqrt{x^2+y^2}}{xy} \ I_1({\sqrt{x^2+y^2}})-\frac{1}{y} I_1(x)-\frac{1}{x} I_1(y)\, .
\end{split}
\ee%

In our analysis, we are interested in the large $x, y$ limit whereby
the last two terms in \eqref{expo} are \emph{exponentially
suppressed} compared to the first term and  may thus be dropped.


\end{document}